\documentclass[aps,prd,preprintnumbers,showpacs,showkeys,nofootinbib,
superscriptaddress,fleqn,floatfix,tightenlines,10pt]{revtex4-2}
\usepackage{amsmath,amsfonts,amssymb,amscd,amsxtra,amsthm}
\usepackage{graphicx}  
\usepackage{epstopdf}
\usepackage{dcolumn}  
\usepackage{bm}          
\usepackage{slashed}
\usepackage{cancel}
\usepackage{float}
\usepackage{mathtools}
\usepackage{amsbsy}
\usepackage{amstext}

\usepackage[utf8]{inputenc} 
\usepackage{booktabs}
\usepackage[normalem]{ulem} 
\usepackage[dvipsnames]{xcolor} 
\usepackage{tabularx}
\usepackage{enumitem}  
\usepackage{array} 
\usepackage{slashed}
\usepackage{tikz}
\usepackage{float}
\usepackage{multirow}
\usepackage{cancel}
\renewcommand\sout{\bgroup \color{red} \ULdepth=-.5ex \ULset}

\makeatletter

\begin{document}  
\title{Quark distribution functions and spin-flavor structures in $N \to 
\Delta$ transitions}
\author{June-Young Kim}
\email[E-mail: ]{jykim@jlab.org}
\affiliation{Theory Center, Jefferson Lab, Newport News, VA 23606, USA}
\begin{abstract}
Transition generalized parton distributions have emerged as a novel tool for studying the quantum chromodynamics~(QCD) structure of resonances. They provide an integrated picture of the transition form factors and the transition parton distribution functions. In this study, we delve into the angular momentum~(AM) properties for the $N\to \Delta$ transition and its decomposition into the orbital angular momentum and the intrinsic spin in the context of the quark distribution functions. First, we explore the spin-flavor structures within the framework of both the overlap representations of the three-quark light-cone wave functions and the large $N_c$ limit of QCD. We then estimate the AM quark distribution functions for the $N\to \Delta$ transition. Our analysis reveals a substantial flavor asymmetry present in both the orbital angular momentum and intrinsic spin components.
\end{abstract}
\date{\today}
\maketitle

\section{Introduction}

The study of nucleon tomography through generalized parton distributions~(GPDs) has been a prominent topic in hadron physics for several decades, offering valuable insights into the internal structure of hadrons; see Refs.~\cite{Polyakov:2002yz, Belitsky:2005qn, Diehl:2003ny} for a review. The GPDs are defined as functions of the skewness $\xi$, the longitudinal momentum fraction of the nucleon carried by the partons $x$, and the squared momentum transfer $t$ between initial and final states. The GPDs parametrize the matrix element of the non-local quark and gluon operators on the light cone and provide an integrated picture of the parton distribution functions (PDFs) and the elastic form factors. In the forward limit, at zero $\xi$ and $t$, the GPDs are reduced to the nucleon PDFs. Furthermore, the Mellin moments of the GPDs are related to the elastic form factors.

Similar to nucleon GPDs, transition GPDs can be introduced as the non-diagonal matrix elements of non-local QCD operators. These transition matrix elements can involve the nucleon and excited states, such as $N \to N^*$ or even $N^* \to N^*$. The study of transition GPDs offers a new avenue to explore the QCD structure and the dynamic properties of resonances. Recently, the analysis of the polarized cross section from the hard exclusive $\pi^{-}\Delta^{++}$ electroproduction off an unpolarized hydrogen target has been performed by the CLAS collaboration~\cite{CLAS:2023akb} and will reveal the $N\to \Delta$ transition GPDs. Theoretical studies have also been performed on the exclusive electroproduction of the $\pi-\Delta$ final states~\cite{Kroll:2022roq} and the deeply virtual Compton process $e^{-}N \to e^{-} \gamma \pi N$~\cite{Semenov-Tian-Shansky:2023bsy}.

The first Mellin moments of the transition vector and axial vector GPDs at zero momentum transfer $t = 0$ correspond to the isovector components of the anomalous magnetic moments $\kappa^{u-d}_{p\to \Delta^{+}}$ and axial charge $g^{u-d}_{A,p\to \Delta^{+}}$, respectively. In addition, the second Mellin moments, known as the  ``transition angular momentum (AM)" $J^{u-d}_{p\to \Delta^{+}}$ in $N\to \Delta$ transitions, have recently been studied and related to the transition vector GPDs in the large $N_{c}$ limit of QCD in Ref.~\cite{Kim:2023xvw}. In addition, the relation between the quadrupole GPDs $H_{E}$ and $H_{C}$ and the isovector quadrupole EMT form factors has been established, together with mechanical interpretations of the transition energy-momentum tensor (EMT) form factors~\cite{multipole}. The Lorentz structure of the transition matrix element of the EMT current has been studied in Ref.~\cite{Kim:2022bwn}, and the corresponding Lorentz invariant functions have been estimated using the light-cone QCD sum rule~\cite{Ozdem:2022zig}. However, the forward limits of the GPDs are still unknown. The study of the transition PDFs is a crucial aspect of the understanding of the $N\to\Delta$ transitions by nucleon tomography.

Reference~\cite{Kim:2023xvw} discusses the definition of the transition AM on the light cone, which provides a clear and unambiguous two-dimensional mechanical interpretation as opposed to a three-dimensional distribution. This approach was exemplified by the interpretation of the non-diagonal matrix element of the $+$-component of the electromagnetic current as a light-cone charge distribution~\cite{Carlson:2007xd, Tiator:2008kd}. However, the price to pay is that the longitudinal information of the distribution is lost and taken over by the quark distribution functions. Numerous studies have addressed this problem for the nucleon~\cite{Burkardt:2000za, Burkardt:2002hr, Belitsky:2005qn, Miller:2007uy, Lorce:2020onh, Epelbaum:2022fjc, Jaffe:2020ebz}. Therefore, the tomography of the $N\to \Delta$ transition must include the study of its quark distribution function.

In this study, we aim to investigate the quark distribution functions using the three-quark ($3Q$) light-cone wave function (LCWF). The unique properties of the relativistic $3Q$ LCWF for the nucleon and its partial wave structure have been extensively studied~\cite{Dziembowski:1987es, Ji:2002xn, Lorce:2011ni}. Building on this uniqueness, we explore the spin-flavor structure of the AM in terms of the overlap representation of the $3Q$ LCWFs. Furthermore, we address a dynamic aspect concerning the decomposition of the total AM into the intrinsic spin and the orbital angular momentum (OAM). The configuration of the LCWFs is derived from the light-cone chiral quark-soliton model (LC$\chi$QSM). In this model, the baryon is viewed as $N_{c}$ valence quarks bound by the pion mean field, and the corresponding system is boosted to the infinite momentum frame and reconstructed to the LCWFs. It was first developed by Diakonov, Petrov and Polyakov~\cite{Petrov:2002jr, Diakonov:2004as} and then elaborated by Lorcé~\cite{Lorce:2006nq, Lorce:2007fa, Lorce:2007as, Lorce:2011dv}. More recently, the normalization $f_{N}$ of the LCWF was determined in Ref.~\cite{Kim:2021zbz}.

An alternative method to study the quark distribution functions in the $N\to \Delta$ transition is to use the standard chiral quark-soliton model ($\chi$QSM)~\cite{Christov:1995vm, Wakamatsu:1990ud}. This model embodies the spin-flavor symmetry in the large $N_{c}$ limit of QCD~\cite{Gervais:1983wq, Gervais:1984rc, Dashen:1993as, Dashen:1993jt, Dashen:1994qi}. Using this spin-flavor symmetry and having the quark distribution functions for the nucleon as input, one can estimate the corresponding distributions for the $N\to \Delta$ transition and the $\Delta$ baryon. Moreover, a notable advantage of this approach is its explicit inclusion of all contributions from quark-antiquark pairs to the quark distribution functions, which are neglected in the overlap representation of the $3Q$ LCWFs; see Ref.~\cite{Schweitzer:2012hh} for more details. This model has been successful in describing the light quark flavor asymmetry~\cite{Diakonov:1997vc, Diakonov:1996sr}, the Gottfried sum rule~\cite{Blotz:1995tj, Pobylitsa:1998tk}, the transversity distributions~\cite{Kim:1995bq, Schweitzer:2001sr}, and the strangeness in scalar~\cite{Kim:1995hu} and vector~\cite{Silva:2001st} and axial-vector~\cite{Blotz:1993am, Blotz:1994wi} charges. More recently, the quasi-parton distributions~\cite{Son:2019ghf, Son:2022qro} have been studied. Therefore, to obtain more realistic results for the quark distribution functions in the $N\to \Delta$ transition, we take data from the $\chi$QSM on the nucleon quark distribution functions~\cite{Diakonov:1997vc, Diakonov:1996sr}.

The structure of this paper is as follows. In Section~\ref{sec:2}, we begin by introducing the formal definitions of the longitudinally polarized and OAM quark distribution functions for a non-diagonal matrix element, starting from the EMT current. In section~\ref{sec:3}, we study the spin-flavor structures of these quark distribution functions in the context of the overlap of the $3Q$ LCWFs. We evaluate the contributions of the OAM and the intrinsic spin to the total AM in the $N\to\Delta$ transition. In section~\ref{sec:4} we establish the relations between the quark distribution functions for the nucleon, the $\Delta$ baryon, and the $N\to\Delta$ transition in the context of the large $N_{c}$ limit. We also establish a connection between the transition quark distribution functions and the axial vector GPDs. In order to obtain more realistic results for the quark distribution functions, we have taken data from the $\chi$QSM on the quark distribution functions of the nucleon. From this we derive the corresponding distributions for the $\Delta$ baryon and the $N\to\Delta$ transition. Finally, we summarize our results in the concluding section.

\section{Quark distribution functions \label{sec:2}}
In Ref.~\cite{Kim:2023xvw}, the QCD AM and its form factor in the $N\to \Delta$ transition were newly introduced. In this work, we provide the separation of the AM into the OAM and the intrinsic spin, and the $x$-dependent distributions~(quark distribution functions).

Before discussing the definition of the quark distribution functions, we start with the QCD EMT current. The $+i$-components of the energy-momentum tensor are related to the normalizations of the longitudinally polarized $\Delta q$ and OAM $l^{q}$ quark distribution functions~\cite{Leader:2013jra}, where $q$ denotes the quark flavor. According to Ji's decomposition~\cite{Ji:1996nm}, the quark part of the EMT current is expressed as
\begin{align}
\hat{T}^{\mu \nu}_{\mathrm{kin},q}  &= \frac{i}{2} \bar{\psi}_{q} \left( 
 \gamma^{\mu} \overleftrightarrow{D}^{\mu}
   \right) \psi_{q}, 
\end{align}
where $\overleftrightarrow{D}^{\mu}=\overleftrightarrow{\partial}_{\mu}- 2 i g A^{\mu}$ with $\overleftrightarrow{\partial}^{\mu} =
\overrightarrow{\partial}^{\mu}- \overleftarrow{\partial}^{\mu}$. The symmetric and antisymmetric parts of the kinetic EMT current can be represented by the divergence of the spin density and the Belinfante-Rosenfeld EMT current, respectively. These expressions are described in the Refs.~\cite{Leader:2013jra, Lorce:2017wkb}:
\begin{align}
\hat{T}^{[\mu\nu]}_{\mathrm{kin},q} = - \partial_{\alpha}\hat{S}^{\alpha \mu \nu}_{q}, \quad \hat{T}^{\{\mu \nu \}}_{\mathrm{kin}, q} =  2 \hat{T}^{\mu \nu}_{q},
\end{align}
where $\hat{T}^{\mu \nu}_{q}$ is the Belinfante-Rosenfeld EMT current, and $\{ab\}=ab+ba$, $[ab]=ab-ba$. It is important to note that while the isoscalar EMT current is conserved, the isovector component of the EMT current is not conserved. In the symmetric frame on the light cone, known as the generalized Drell-Yan-West (DYW) frame~\cite{Kim:2023xvw} with $\bm{P}_{\perp}=0$ and $\Delta^{+}=0$, the EMT distributions are defined by the Fourier transform of the baryonic matrix element of the EMT current~\cite{Kim:2023xvw}
\begin{align}
T^{+ i}_{q} (b | B' B) = \int \frac{d^{2}\Delta}{(2\pi)^{2}} e^{-i \bm{\Delta}\cdot \bm{b}} \langle B'(P^{+},\bm{\Delta}/2, S'_{3})| \hat{T}^{+i}_{q}  | B (P^{+},-\bm{\Delta}/2, S_{3}) \rangle,
\label{eq:def}
\end{align}
where the initial and final baryons are denoted by $B$ and $B'$, respectively. The light-cone vectors are defined by $v^{\pm}= (v^{0} \pm v^{3})\sqrt{2}$. The average and the difference between the incoming~$(p)$ and outgoing~$(p')$ baryon momenta are  given by $P= (p'+p)/2$ and $\Delta= p'-p$, respectively. The impact parameter and the momentum transfer lie in the transverse plane, i.e., $\bm{b}=\{b_{x}, b_{y}\}$ and $\bm{\Delta}=\{\Delta_{x}, \Delta_{y}\}$. While $S'_{3}$ and $S_{3}$ denote the spin projections of the initial and final baryons, respectively, the baryon spin states in Eq.~\eqref{eq:def} are chosen as light-front helicity states. 

The QCD AM can be separated into two components: the OAM denoted by $l^{q}$, which is associated with the kinetic EMT current $T^{+i}_{\mathrm{kin},q}$, and the intrinsic spin represented by $\Delta q$, which corresponds to the antisymmetric part of the kinetic EMT current $T^{[+i]}_{\mathrm{kin},q}$. The QCD AM is then defined by
\begin{align}
2S^{z}(S'_{3},S_{3}) J^{q}_{B' \to B}& = 2S^{z}(S'_{3},S_{3}) [l^{q} + \Delta q ]_{B' \to B} = \frac{1}{2P^{+}} \int d^{2}\bm{b} \, [\bm{b} \times \bm{T}^{+ T}_{q} (b | B' B)]^{z}, 
\end{align}
with the matrix element of the generalized spin vector
\begin{align}
S^{z}(S'_{3},S_{3})  = \sqrt{S(S+1)}\sqrt{ \frac{2S+1}{2S'+1}} C^{S' S'_{3}}_{S S_{3} 1 0},
\label{eq:spin}
\end{align}
where $S'$ and $S$ represent the spins of the initial and final states, respectively, and $C^{S' S'_{3}}_{S S_{3} 1 0}$ corresponds to the SU(2) Clebsch-Gordan coefficient. The equation~\eqref{eq:spin} holds not only for the diagonal matrix element $|S'-S|=0$, but also for the non-diagonal transition $|S'-S|=1$. The standard spin vector is recovered by setting $S'=S$.  By inserting the different initial and final spin quantum numbers, one can define the transition angular momentum~\cite{Kim:2023xvw}. 


Similarly, we can introduce and interpret the longitudinally polarized and OAM quark distribution functions for the $N \to \Delta$ transition as the forward limits of the transition GPDs. While the forward limit of the twist-2 nucleon axial-vector GPDs~\cite{Polyakov:2002yz, Belitsky:2005qn, Diehl:2003ny} is related to the longitudinally polarized quark distribution, the OAM quark distribution is related to the twist-3 nucleon GPDs~\cite{Penttinen:2000dg, Ji:2012ba, Hatta:2012cs}. Similar relations can be found for the $N \to \Delta$ transition, which will be discussed in section~\ref{sec:4}. Neglecting the gauge-field degrees of freedom, the longitudinally polarized $\Delta q_{B\to B'}(x)$ and OAM $l^{q}_{B'\to B}(x)$ quark distributions are formally given by (see Refs.~\cite{Polyakov:2002yz, Belitsky:2005qn, Diehl:2003ny, Leader:2013jra} for a review)
\begin{align}
& 2 S^{z}(S'_{3},S_{3}) \Delta q_{B\to B'}(x) = \frac{1}{2} \int \frac{dz^{-}}{2\pi} e^{i x P^{+} z^{-}} \langle B' (P^{+}, \bm{0}, S'_{3}) | \bar{\psi}_{q}\left(-\frac{z}{2}\right) \gamma^{+}\gamma^{5} \psi_{q}\left(\frac{z}{2}\right)  |  B (P^{+}, \bm{0}, S_{3}) \rangle \bigg{|}_{z^{+}=z_{\perp}=0}, \cr
& 2 S^{z}(S'_{3},S_{3}) l^{q}_{B'\to B}(x) = \int d^{2}\bm{k}_{\perp}  i (\bm{k}_{\perp}\times \bm{\nabla}_{\Delta})_{z} \frac{1}{2} \int \frac{dz^{-} d^{2}\bm{z}_{\perp}}{(2\pi)^{3}} e^{i (x P^{+} z^{-} - \bm{k}_{\perp}\cdot \bm{z}_{\perp})} \cr
&\hspace{4.5cm} \times \langle B'(P^{+},\bm{\Delta}/2, S'_{3}) | \bar{\psi}_{q}\left(-\frac{z}{2}\right) \gamma^{+} \psi_{q}\left(\frac{z}{2}\right)  |  B(P^{+},-\bm{\Delta}/2, S_{3})  \rangle \bigg{|}_{z^{+}=0,\bm{\Delta}=0},
\label{eq:def_pdf}
\end{align}
where $x$ is the longitudinal momentum fraction of a baryon carried by quarks. Thus, the total AM quark distribution function is given by
\begin{align}
J^{q}_{B\to B'}(x) :=  l^{q}_{B\to B'}(x) +\Delta q_{B\to B'}(x).
\end{align}


\section{Light-cone chiral quark-soliton model \label{sec:3}}
The chiral quark-soliton model~($\chi$QSM) is a pion mean-field theory, motivated by the large $N_{c}$ limit of QCD~\cite{Witten:1979kh}. The pion field is created by the presence of valence quarks, and the valence quarks are in turn influenced by the pion field. This self-consistent interaction leads to the formation of a baryon; see Refs.~\cite{Christov:1995vm, Wakamatsu:1990ud} for s review. The light-cone chiral quark soliton model~(LC$\chi$QSM) is a version of the $\chi$QSM that is formulated on the light cone~\cite{Petrov:2002jr, Diakonov:2004as}. This allows for a more simpler treatment of the relativistic dynamics of the model; see also Ref.~\cite{Schweitzer:2012hh} for a study of the LCWFs at large $N_c$.

\subsection{Light-cone wave functions}
In this model, the baryon wave function appears as the sum of the discrete-level wave function\footnote{In fact, the configuration of the discrete-level wave function is slightly distorted by the vacuum polarization effects. Since these contributions are negligible, we omit them in this work. Detailed estimates of such contributions can be found in Ref.~\cite{Lorce:2011dv}.} $F^{j\sigma}$ and the infinite tower of quark-antiquark pair wave functions. While higher Fock states, such as $5Q$ and $7Q$, can be generated by the pair wave functions, we limit our investigation to the $3Q$ configuration. This choice is motivated by the observation that the inclusion of higher Fock states would introduce corrections to the observables that are typically no larger than 20\% in many theoretical approaches. Therefore, we will consider only the discrete-level wave function
\begin{align}
  &F^{j\sigma}(x,\bm{k}_{\perp}) =  \left(\begin{array}{c c } k_{L}
 f_{\perp}(x,|\bm{k}_{\perp}|) & f_{\parallel}(x,|\bm{k}_{\perp}|) \\ -f_{\parallel}(x,|\bm{k}_{\perp}|) & k_{R}f_{\perp}(x,|\bm{k}_{\perp}|) \end{array}\right)^{j\sigma}
\bigg{|}_{k_{z}=xM_{N}-E_{\mathrm{lev}}},
\end{align}
where $j$ is the quark isospin and $\sigma$ is the light-cone helicity with $k_{R,L}=k_{x}\pm i k_{y}$ and the nucleon mass $M_{N}$. Here, the longitudinal momentum fraction of the baryon carried by the quark is denoted by $x$, and the transverse momentum of the quark is given by $\bm{k}_{\perp}$. The two independent functions $f_{\parallel}(x,|\bm{k}_{\perp}|)$ and $f_{\perp}(x,|\bm{k}_{\perp}|)$ are written as
\begin{align}
  & f_{\parallel}(x,|\bm{k}_{\perp}|) 
=\sqrt{\frac{M_{N}}{2\pi}}\left(  h(k) + \frac{k_{z}j(k)}{|\bm{k}|}\right), \ \ \  f_{\perp}(x,|\bm{k}_{\perp}|) =  \sqrt{\frac{M_{N}}{2\pi}} \frac{j(k)}{|\bm{k}|}.
\end{align}
$h$ and $j$ are the upper and lower components of the Dirac spinor in the presence of the chiral fields. $E_{\mathrm{lev}}$ is the discrete-level quark eigenenergy bounded by the pion mean field. Note that in the nonrelativistic~(NR) limit the diagonal part of $F^{j\sigma}$ is dropped, i.e,
\begin{align}
  & f^{\mathrm{NR}}_{\parallel}(x,|\bm{k}_{\perp}|) 
=\sqrt{\frac{M_{N}}{2\pi}}  h(k) , \ \ \  f^{\mathrm{NR}}_{\perp}(x,|\bm{k}_{\perp}|) =  0.
\end{align}
After summing over the discrete-level energy $E_{\mathrm{lev}}\sim 0.2~\mathrm{GeV}$ and the Dirac continuum spectra, one finds the classical nucleon mass $M_{N}\sim 1.207~\mathrm{GeV}$.

In the limit of large $N_{c}$, the baryon wave function can be completely factorized in color space, so that the LCWFs are given by the product of the $N_{c}$ discrete-level wave functions~(see Refs.~\cite{Petrov:2002jr, Diakonov:2004as, Diakonov:2005ib} for the details):
\begin{align}
|B (P,S_{3})\rangle &=   T(B)^{f_1f_2f_3}_{j_{1}j_{2}j_{3},k}    \frac{c_{0}}{\sqrt{P_{z}}}   \int[d\bm{k}]\int[d x] F^{j_1\sigma_1}(\bm{k}_1)F^{j_2\sigma_2}(\bm{k}_2)F^{j_3\sigma_3}(\bm{k}_3)  \cr
&\times \frac{\epsilon^{\alpha_{1}\alpha_{2}\alpha_{3}}}{\sqrt{N_{c}!}} a^{\dagger}_{\alpha_1 f_1 \sigma_1}(\bm{p}_1) a^{\dagger}_{\alpha_2 f_2 \sigma_2}(\bm{p}_2) a^{\dagger}_{\alpha_3 f_3 \sigma_3}(\bm{p}_3) | 0  \rangle,
\label{eq:cov_WF}
\end{align}
with the relative transverse momenta of the quarks $\bm{k}_{i}=(x_{i},\bm{k}_{i\perp})$ and the physical transverse momenta $\bm{p}_{i}=(x_{i},\bm{k}_{i\perp}+x_{i} \bm{P}_{\perp})$. The index $\alpha_{i}$ denotes the color index, and the light-cone wave function is antisymmetric in the color space. The indices $f_{i}$ and  $\sigma_{i}$ stand for the quark flavors and the light-cone helicity, respectively. $\bm{P}= (\bm{P}_{\perp},P_{z})$ and $c_{0}$ designate the baryon momentum and the normalization constant of the light-cone wave function, respectively. The quark creation operator, denoted as $a^{\dagger}$, follows the anticommutation relations
\begin{align}
\{a^{\dagger}_{a}(\bm{p}_{\perp}) , a_{a'}(\bm{p}')\} = \delta_{a' a} \delta(x-x') (2\pi)^{2} \delta^{(2)} (\bm{p}_{\perp}' - \bm{p}_{\perp}).
\end{align}
with $a=\{\alpha, f, \sigma\}$ and $a'=\{\alpha', f', \sigma'\}$. In order to ensure a well-defined momentum for the baryon state, we take into account the translational zero mode. This leads to the conservation of momentum in the baryon state. Consequently, the integration measure for the $3Q$ state can be expressed as follows:
\begin{align}
  \int [dx] = \int dx_1 dx_2 dx_3\, \delta\left(\sum^{3}_{l=1}
  x_{l}-1   \right),    \quad  \int [d\bm{k}] = \int \left( \prod_{i=1}^3
  \frac{d^{2}\bm{k}_{i\perp}}{(2\pi)^{2}} \right)
  (2\pi)^{2}\delta^{(2)}\left(\sum^{3}_{l=1}
  \bm{k}_{l\perp}\right).
\label{eq:Imeasure}
\end{align}
To include the spin and isospin quantum numbers of the baryon, we also consider the rotational zero mode. Each of the discrete-level quarks undergoes a rotation given by the matrix $R^{f}_{j}$ and is projected onto the spin-flavor baryon state $B^*(R)$ by integrating over $R$. We denote this group integral by the following shorthand:
\begin{align}
  &T(B)^{f_{1}f_{2}f_{3}}_{j_{1}j_{2}j_{3},k} :=
  \int dR B^{*}_{k}(R)R^{f_{1}}_{j_{1}} R^{f_{2}}_{j_{2}} R^{f_{3}}_{j_{3}},
\end{align}
where the spin-flavor baryon states are given by
\begin{align}
p^{*}_{\uparrow}=\sqrt{8}R^{\dagger 2}_{1}R^{3}_{3}, \quad \Delta^{+*}_{\uparrow} = \sqrt{10} (R^{\dagger2}_{1}R^{\dagger2}_{1}R^{\dagger1}_{2} + R^{\dagger2}_{2}R^{\dagger2}_{1}R^{\dagger1}_{1}).
\end{align}
The spin polarization of the baryon is denoted by $k=\uparrow, \downarrow= 1,2$. The prefactors of the spin-flavor baryon states are determined through the following normalization:
\begin{align}
\int dR B^{*}_{k}(R) B^{k}(R) = 1.
\end{align}
We refer to Refs.~\cite{Diakonov:2005ib, Lorce:2006nq,Lorce:2007as, Lorce:2007xax} for the details of the group integrals. 

Note that in the large $N_{c}$ limit we set the masses of the nucleon and the $\Delta$ baryon to be equal, i.e., $M_{N}=M_{\Delta} \sim \mathcal{O} (N_{c})$. To include higher-order corrections beyond the leading $N_{c}$ approximation, the integration measure must be adjusted. This modification accounts for the kinematical subleading corrections in $N_{c}$ and has been discussed in Ref.~\cite{Cedric:2007vc}.

\subsection{Normalization of the LCWFs}
In this section, we focus on the parameterization of the overlap integrals of the LCWFs. This parameterization allows us to determine the normalization of the LCWFs. Specifically, the normalization factor $c_{0}= \sqrt{(2\pi)/\mathcal{N}^{(3)}}$ of the baryon LCWF can be obtained by evaluating the contraction of the creation and annihilation operators, which is written as
\begin{align}
\mathcal{N}^{(3)}(B) &= 6 T(B)^{f_1f_2f_3}_{j_1j_2j_3,k} T(B)_{f_1f_2f_3}^{j'_1j'_2j'_3,k}  \int [d\bm{k}]
    F^{j_1\sigma_1}(\bm{k}_1)F^{j_2\sigma_2}(\bm{k}_{2})F^{j_3\sigma_3}(\bm{k}_{3}) F^{\dagger}_{j'_1\sigma_1}(\bm{k}_1) F^{\dagger}_{j'_2\sigma_2}(\bm{k}_2) F^{\dagger}_{j'_3\sigma_3}(\bm{k}_3).
\label{eq:3Qnor}
\end{align}
By performing the group integrals for both the proton and the $\Delta$ baryon, we determine their respective normalizations
\begin{align}
\mathcal{N}^{(3)}(p_{\uparrow})=\frac{3}{2} \alpha^{V}, \quad \mathcal{N}^{(3)}(\Delta^{+}_{\uparrow})=\frac{3}{5} \alpha^{V}, 
\end{align}
where it is convenient to define the $\alpha^{V}$ in terms of the quark distributions $\Phi^{V}(x)$
\begin{align}
\alpha^{I=V}:=\int dx \, \Phi^{I=V}(x) &= \int[dx] \int [d\bm{k}] D^{I=V}(x,\bm{k}_1,\bm{k}_2,\bm{k}_3),
\label{eq:valence_prob}
\end{align}
with
\begin{align}
D^{I=V}=\delta(x-x_{1}) \bigg{[}f^{2}_{\parallel}(\bm{k}_1)+k_{1R}k_{1L}f^{2}_{\perp}(\bm{k}_1)\bigg{]}\bigg{[}f^{2}_{\parallel}(\bm{k}_2)+k_{2R}k_{2L}f^{2}_{\perp}(\bm{k}_2)\bigg{]}\bigg{[}f^{2}_{\parallel}(\bm{k}_3)+k_{3R}k_{3L}f^{2}_{\perp}(\bm{k}_3)\bigg{]}. 
\end{align}
Note that the normalizations of the discrete-level wave functions $f_{\perp}$ and $f_{\parallel}$ are arbitrary. For convenience, we choose them to have a normalization parameter $\alpha^{V}$ equal to $1$.

\subsection{Overlap integrals}
We are now in a position to evaluate the quark distribution functions of the AM using the LCWFs. The overlap representations for the AM operators are parameterized by the five quark distributions, and they are labeled by $I=A,L,L_1,L_{2},L_{3}$. $D^{I=A,L,L_1,L_2,L_{3}}(x,\bm{k}_1,\bm{k}_2,\bm{k}_3)$ is then defined as 
\begin{align}
&D^{A}= \delta(x-x_{1})\bigg{[}f^{2}_{\parallel}(\bm{k}_1)-k_{1R}k_{1L}f^{2}_{\perp}(\bm{k}_1)\bigg{]}\bigg{[}f^{2}_{\parallel}(\bm{k}_2)+k_{2R}k_{2L}f^{2}_{\perp}(\bm{k}_2)\bigg{]}\bigg{[}f^{2}_{\parallel}(\bm{k}_3)+k_{3R}k_{3L}f^{2}_{\perp}(\bm{k}_3)\bigg{]}, \cr
&D^{L}= \delta(x-x_{1})\bigg{[}x_{1}k_{1R}k_{1L}f^{2}_{\perp}(\bm{k}_1)\bigg{]}\bigg{[}f^{2}_{\parallel}(\bm{k}_2)+k_{2R}k_{2L}f^{2}_{\perp}(\bm{k}_2)\bigg{]}\bigg{[}f^{2}_{\parallel}(\bm{k}_3)+k_{3R}k_{3L}f^{2}_{\perp}(\bm{k}_3)\bigg{]}, \cr
&D^{L_1}= \delta(x-x_{1})\bigg{[}(1-x_{1})k_{1R}k_{1L}f^{2}_{\perp}(\bm{k}_1)\bigg{]}\bigg{[}f^{2}_{\parallel}(\bm{k}_2)+k_{2R}k_{2L}f^{2}_{\perp}(\bm{k}_2)\bigg{]}\bigg{[}f^{2}_{\parallel}(\bm{k}_3)+k_{3R}k_{3L}f^{2}_{\perp}(\bm{k}_3)\bigg{]}, \cr
&D^{L_2}=\delta(x-x_{2})\bigg{[}\frac{1}{2} (k_{1R}k_{2L}+k_{1L}k_{2R}) x_{1}f^{2}_{\perp}(\bm{k}_1)\bigg{]}\bigg{[}f^{2}_{\parallel}(\bm{k}_2)+k_{2R}k_{2L}f^{2}_{\perp}(\bm{k}_2)\bigg{]}\bigg{[}f^{2}_{\parallel}(\bm{k}_3)+k_{3R}k_{3L}f^{2}_{\perp}(\bm{k}_3)\bigg{]}, \cr
&D^{L_3}=\delta(x-x_{3})\bigg{[}\frac{1}{2} (k_{1R}k_{3L}+k_{1L}k_{3R}) x_{1}f^{2}_{\perp}(\bm{k}_1)\bigg{]}\bigg{[}f^{2}_{\parallel}(\bm{k}_2)+k_{2R}k_{2L}f^{2}_{\perp}(\bm{k}_2)\bigg{]}\bigg{[}f^{2}_{\parallel}(\bm{k}_3)+k_{3R}k_{3L}f^{2}_{\perp}(\bm{k}_3)\bigg{]}.
\label{eq:valence_prob_func}
\end{align}
Using the quark distributions given in Eq.~\eqref{eq:valence_prob_func}, we obtained the AM quark distribution functions:
\begin{itemize}
\item for the proton (for the neutron ($u\leftrightarrow d$))
\begin{align}
&l^{u}_{p}(x)= \frac{4}{3}\Phi^{L_1}(x) - \frac{1}{3}\Phi^{L_2}(x)- \frac{1}{3}\Phi^{L_3}(x), \quad l^{d}_{p}(x)= -\frac{1}{3}\Phi^{L_1}(x) - \frac{2}{3}\Phi^{L_2}(x)- \frac{2}{3}\Phi^{L_3}(x), \cr
&l^{u+d}_{p}(x) =  \Phi^{L_1}(x) - \Phi^{L_2}(x)- \Phi^{L_3}(x), \quad l^{u-d}_{p}(x) =  \frac{5}{3}\Phi^{L_1}(x) +\frac{1}{3} \Phi^{L_2}(x) +\frac{1}{3} \Phi^{L_3}(x),  \cr
&\Delta u_{p}(x)= \frac{2}{3}\Phi^{A}(x), \quad \Delta d_{p}(x)= -\frac{1}{6}\Phi^{A}(x), \cr
&\left[\Delta u(x) + \Delta d(x)\right]_{p} =  \frac{1}{2}\Phi^{A}(x), \quad \left[\Delta u(x) - \Delta d(x)\right]_{p} = \frac{5}{6}\Phi^{A}(x).
\end{align}
\item for the $\Delta^{+}$ baryon
\begin{align}
&l^{u}_{\Delta^{+}}(x)= \frac{2}{3}\left(\Phi^{L_1}(x) - \Phi^{L_2}(x)-\Phi^{L_3}(x)\right), \quad l^{d}_{\Delta^{+}}(x)= \frac{1}{3}\left(\Phi^{L_1}(x) - \Phi^{L_2}(x)-\Phi^{L_3}(x)\right), \cr
&l^{u+d}_{\Delta^+}(x) =  \Phi^{L_1}(x) - \Phi^{L_2}(x)- \Phi^{L_3}(x), \quad l^{u-d}_{\Delta^+}(x) =  \frac{1}{3}\left(\Phi^{L_1}(x) - \Phi^{L_2}(x)-\Phi^{L_3}(x)\right), \cr
&\Delta u_{\Delta^{+}}(x)= \frac{1}{3}\Phi^{A}(x), \quad \Delta d_{\Delta^{+}}(x)= \frac{1}{6}\Phi^{A}(x), \cr
&\left[\Delta u(x) + \Delta d(x)\right]_{\Delta^{+}} =  \frac{1}{2}\Phi^{A}(x), \quad \left[\Delta u(x) - \Delta d(x)\right]_{\Delta^{+}} = \frac{1}{6}\Phi^{A}(x). 
\end{align}
\item for the $p \to \Delta^{+}$ transition
\begin{align}
&l^{u}_{p\to \Delta^{+}}(x)= -\frac{\sqrt{2}}{3}\left(2\Phi^{L_1}(x) - \Phi^{L_2}(x)-\Phi^{L_3}(x)\right), \quad l^{d}_{p\to \Delta^{+}}(x)= \frac{\sqrt{2}}{3}\left(2\Phi^{L_1}(x) - \Phi^{L_2}(x)-\Phi^{L_3}(x)\right), \cr
&l^{u+d}_{p\to \Delta^{+}}(x) =  0, \quad l^{u-d}_{p\to \Delta^{+}}(x) =  -\frac{2\sqrt{2}}{3}\left(2\Phi^{L_1}(x) - \Phi^{L_2}(x)-\Phi^{L_3}(x)\right), \cr
&\Delta u_{p\to\Delta^{+}}(x)= -\frac{\sqrt{2}}{3}\Phi^{A}(x), \quad \Delta d_{p\to\Delta^{+}}(x)= \frac{\sqrt{2}}{3}\Phi^{A}(x), \cr
&\left[\Delta u(x) + \Delta d(x)\right]_{p\to\Delta^{+}} =  0, \quad \left[\Delta u(x) - \Delta d(x)\right]_{p\to\Delta^{+}} = -\frac{2\sqrt{2}}{3}\Phi^{A}(x). 
\end{align}
\end{itemize}
Based on the spin-flavor structures of the proton~($p$), the neutron~($n$), the $\Delta^{+}$ baryon and the $p\to \Delta^+$ transition, we obtain intriguing relations for the OAM quark distribution functions
\begin{align}
&l^{u}_{p\to \Delta^{+}}(x) = -\frac{\sqrt{2}}{3} \bigg{(}l^{u}_{\Delta^{+}}(x) + l^{u}_{p}(x)\bigg{)} =- \frac{\sqrt{2}}{9} \bigg{(}2l^{u}_{n}(x) + 5l^{u}_{p}(x)\bigg{)}, \cr
&l^{d}_{p\to \Delta^{+}}(x) = \frac{\sqrt{2}}{3} \bigg{(}5l^{d}_{\Delta^{+}}(x) - l^{d}_{p}(x)\bigg{)} = \frac{\sqrt{2}}{9} \bigg{(}5l^{d}_{n}(x) + 2l^{d}_{p}(x)\bigg{)}, 
\end{align}
and for the longitudinally polarized quark distribution functions
\begin{align}
\Delta u_{p\to \Delta^{+}}(x)  = -\frac{1}{\sqrt{2}}  \Delta u_{p}(x)=  -\sqrt{2} \Delta u_{\Delta^{+}}(x), \cr
\Delta d_{p\to \Delta^{+}}(x)  = -2\sqrt{2}  \Delta d_{p}(x)=  2\sqrt{2} \Delta d_{\Delta^{+}}(x).
\end{align}

By integrating the quark distribution functions over the variable $x$, we have obtained the values for the intrinsic spin (also known as the axial charge $\Delta q_{B' \to B} = g^{q}_{A,B'\to B}/2$) and the OAM
\begin{align}
\int dx \,  \Delta q_{B\to B'}(x) = \Delta q_{B\to B'}, \quad  \int dx \, l^{q}_{B\to B'}(x) = l^{q}_{B\to B'}.
\end{align}
From Eq.~\eqref{eq:valence_prob_func}, one can easily see that the integral of the total AM quark distribution functions $J^{u+d}_{p,\Delta^{+}}(x)$ over $x$ is properly normalized to the baryon spin
\begin{align}
\int dx \, J^{u+d}_{p,\Delta^{+}}(x) = \int dx \, \left[\frac{1}{2} \Phi^{A}(x) + \Phi^{L_1}(x) - \Phi^{L_2}(x)- \Phi^{L_3}(x) \right] = \frac{1}{2}.
\label{eq:sumrule}
\end{align}
Here we used the relation $\bm{k}_{1}+\bm{k}_{2}+\bm{k}_{3}=0$. It is obvious that the isoscalar transition AM $J^{u+d}_{p\to\Delta^{+}}(x)$ is equal to zero. It is worth noting that in the nonrelativistic limit, all OAM quark distribution functions become zero, while the longitudinally polarized quark distribution functions become equivalent to the total AM quark distribution functions:
\begin{align}
\Delta q_{B\to B', \mathrm{NR}}(x) = J^{q}_{B\to B', \mathrm{NR}}(x), \quad l^{q}_{B\to B', \mathrm{NR}}(x)=0.
\end{align}

\subsection{Numerical results}
To estimate the AM quark distribution functions and their normalizations, we use the explicit quark wave functions $f_{\perp}$ and $f_{\parallel}$, where the values of the dynamical parameters are taken from Refs.~\cite{Petrov:2002jr, Diakonov:2004as, Diakonov:2005ib, Lorce:2006nq}. We will provide not only the quark distribution functions for the $N\to \Delta$ transition, but also how many fractions of the intrinsic spin and the OAM contribute to the $N\to \Delta$ transition AM.

\begin{figure}[htp]
\includegraphics[scale=0.28]{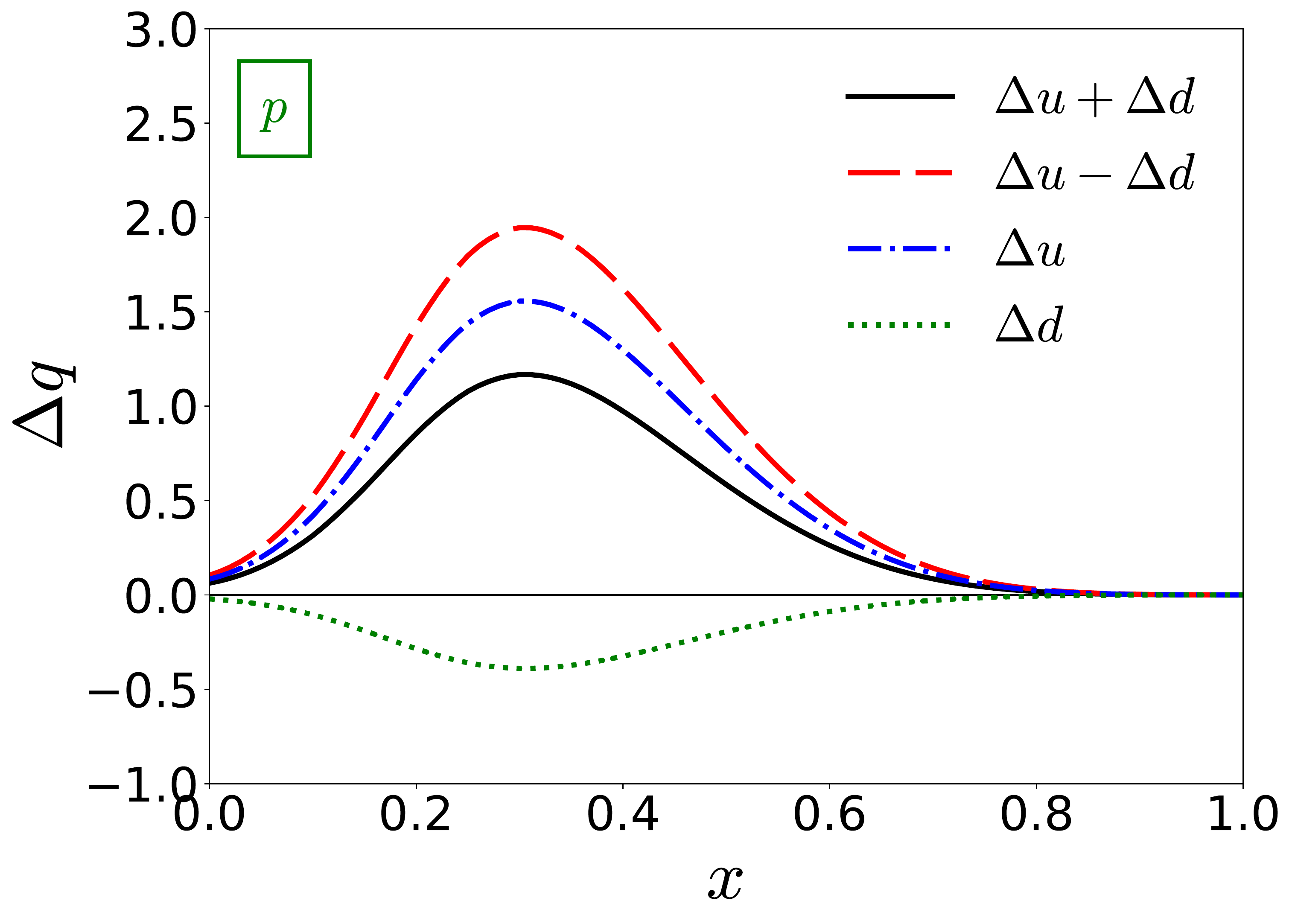}
\includegraphics[scale=0.28]{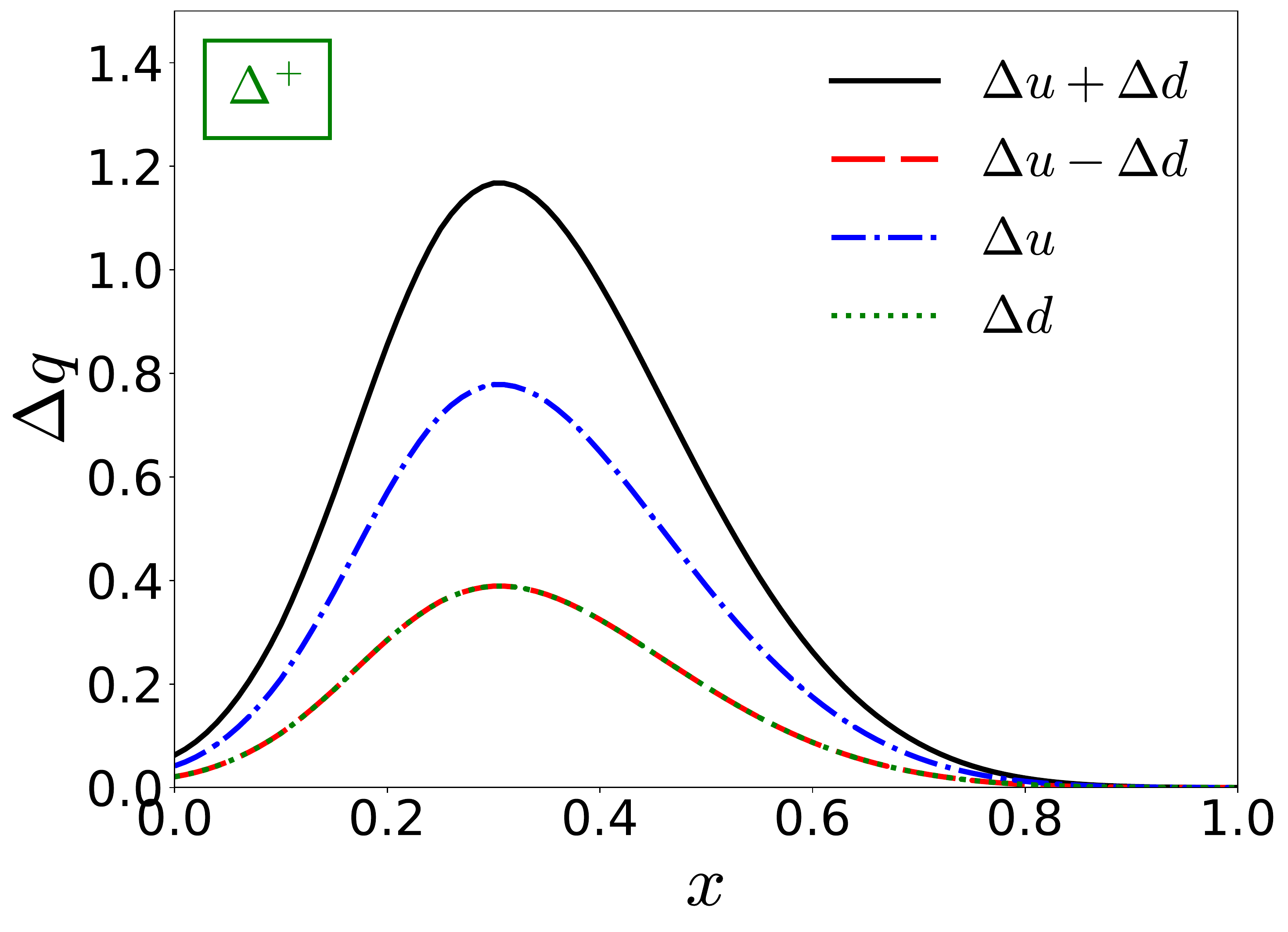}
\includegraphics[scale=0.28]{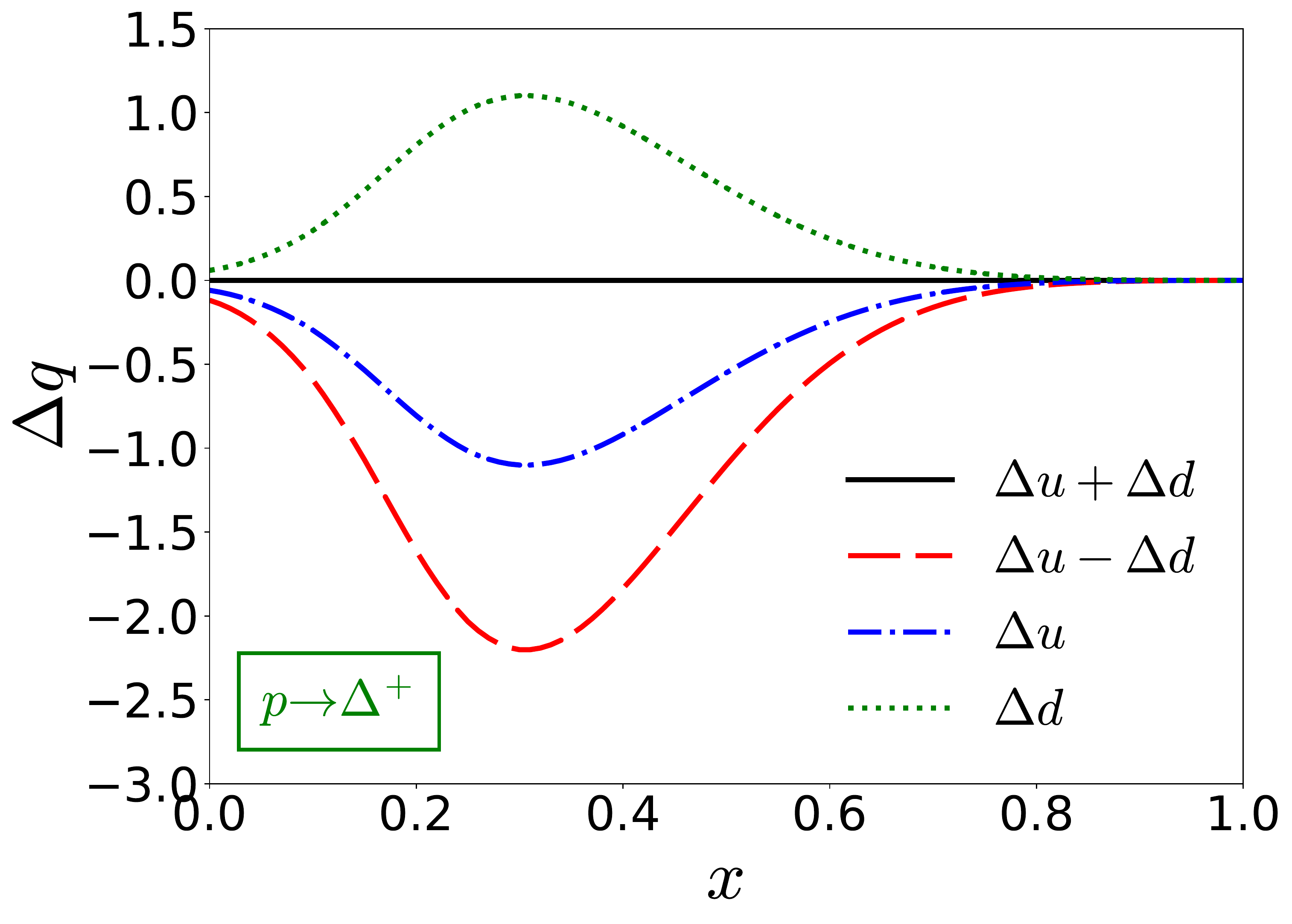}
\caption{Longitudinally polarized quark distribution functions for the proton~(upper left panel), $\Delta$ baryon~(upper right panel), and $p\to \Delta^{+}$ transition~(lower panel). The solid~(black), dashed~(red), dot-dashed~(blue), and dotted~(green) lines represent the $\Delta u+ \Delta d$, $\Delta u- \Delta d$, $\Delta u$, and $\Delta d$ contributions, respectively.}
\label{fig:1}
\end{figure}
In Fig.~\ref{fig:1} we first examined the longitudinally polarized quark distribution functions for the proton, the $\Delta$ baryon, and the $N\to \Delta$ transition. These distribution functions are parameterized with respect to the single quark distribution $\Phi_{A}$, which is normalized as follows:
\begin{align}
\int dx \, \Phi^{A}(x)=\alpha^{A}=0.861, \quad \int dx \, \Phi^{A}_{\mathrm{NR}}(x)= \alpha^{A}_{\mathrm{NR}}=1,
\end{align}
where we have reproduced the numerical values given in Ref.~\cite{Lorce:2007as}.  It is observed that the $\Delta u$ and $\Delta d$ values for the proton have opposite signs, with $\Delta u$ being positive and $\Delta d$ being negative. However, for the $\Delta^{+}$ baryon, both $\Delta u$ and $\Delta d$ have positive signs. Interestingly, in the case of the proton, the isovector component of the axial charge is significantly larger than the isoscalar component. Conversely, for the $\Delta$ baryon, this relation is reversed. Turning to the quark distribution functions for the $p \to \Delta^{+}$ transition, they are naturally induced from the group relation. While the isoscalar quark distribution functions in the $p \to \Delta^{+}$ transition are zero, a substantial asymmetry between the light valence quarks is obtained. The normalizations of these distribution functions are summarized in Table~\ref{tab:1}. Consistent with the large $N_{c}$ analysis~\cite{Kim:2023xvw}, it is noteworthy that the flavor asymmetries in the intrinsic spin $[\Delta u - \Delta d]_{p\to \Delta^{+}} = -0.812$ and in the total AM $J^{u-d}_{p\to \Delta^{+}} = -0.887$ are estimated to be substantial. Note that the sign difference for $J^{u-d}_{p\to \Delta^{+}}$ compared to Ref.~\cite{Kim:2023xvw} might depend on the choice of the phases of the baryon states.

\begin{figure}[htp]
\includegraphics[scale=0.28]{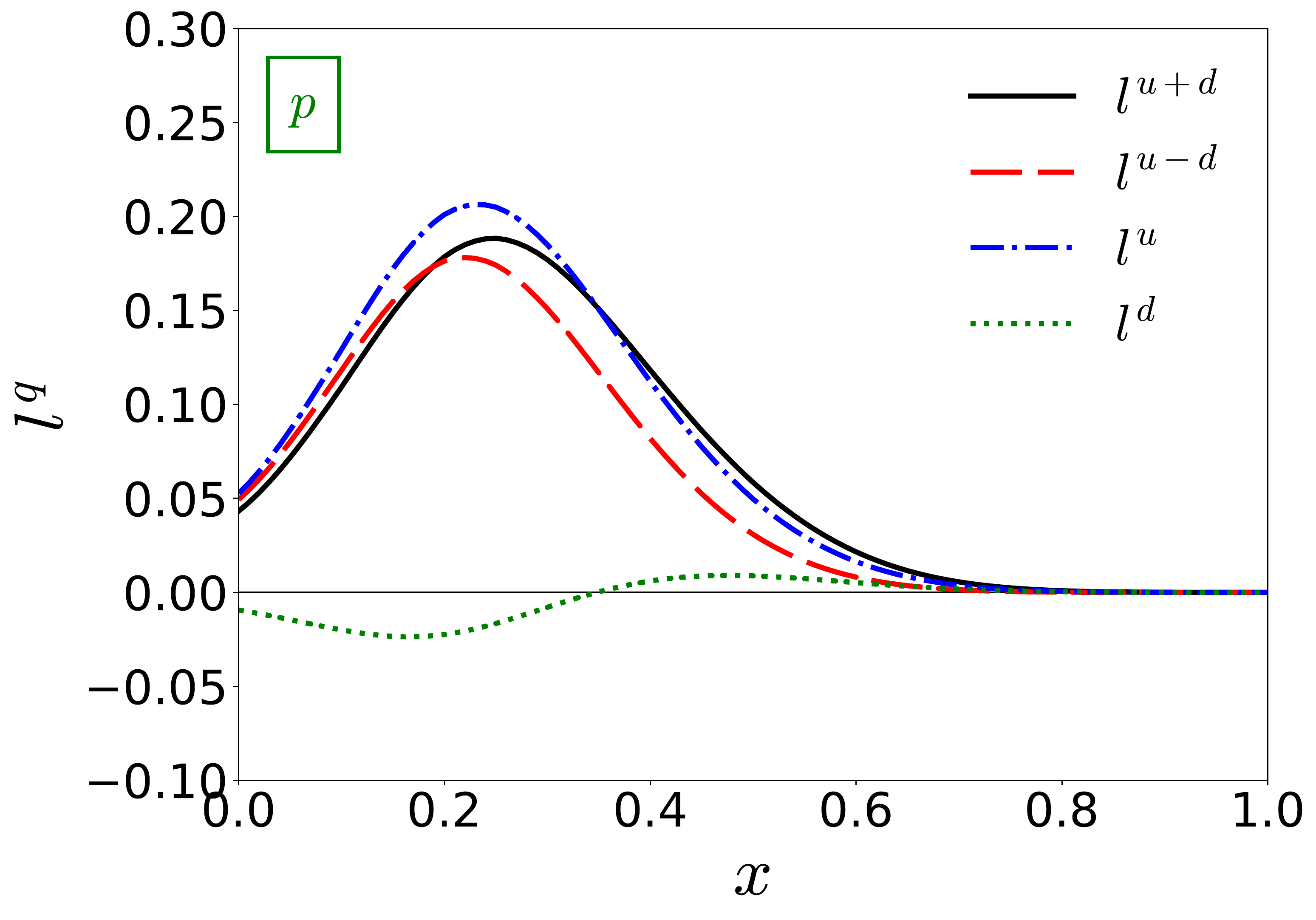}
\includegraphics[scale=0.28]{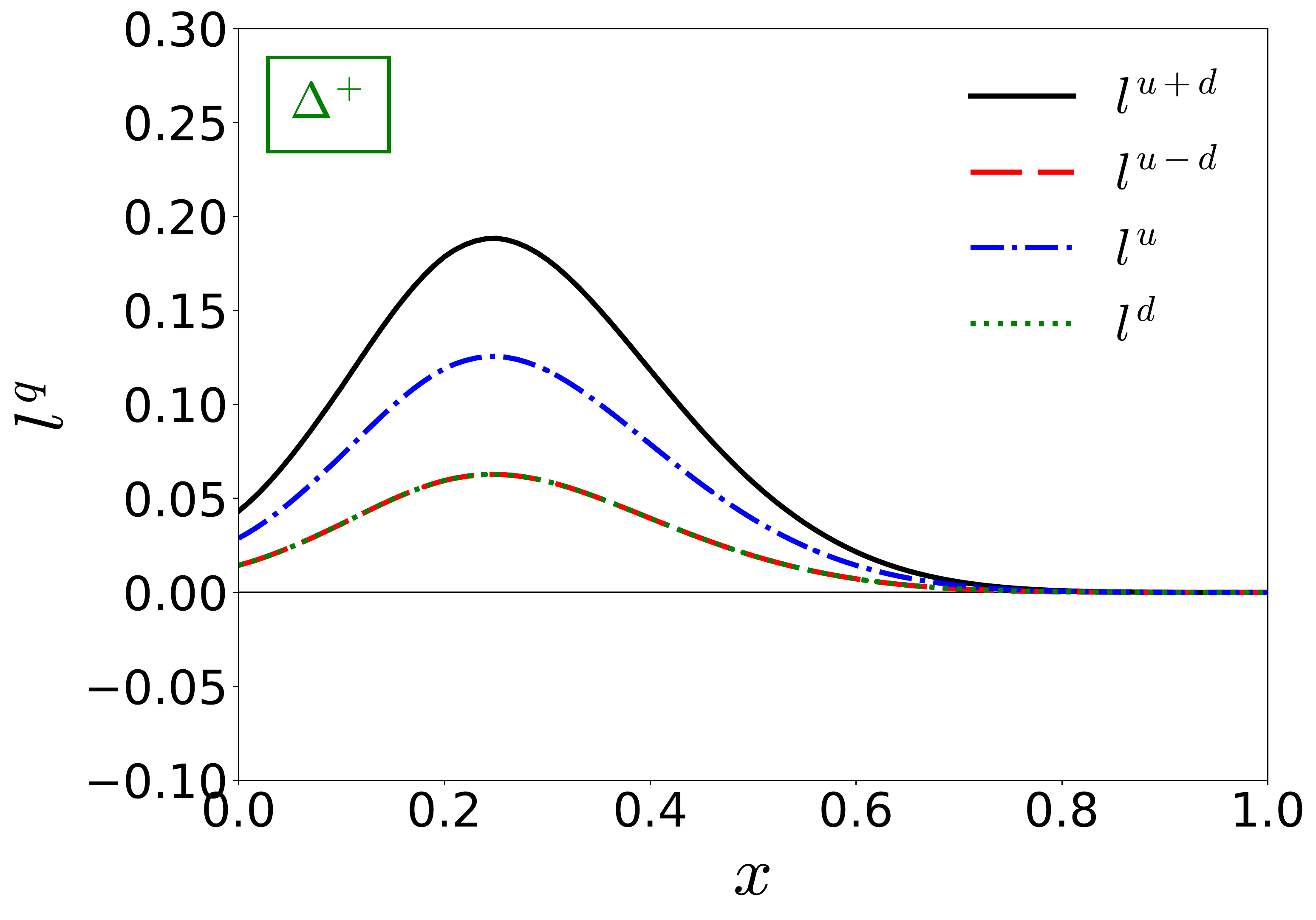}
\includegraphics[scale=0.28]{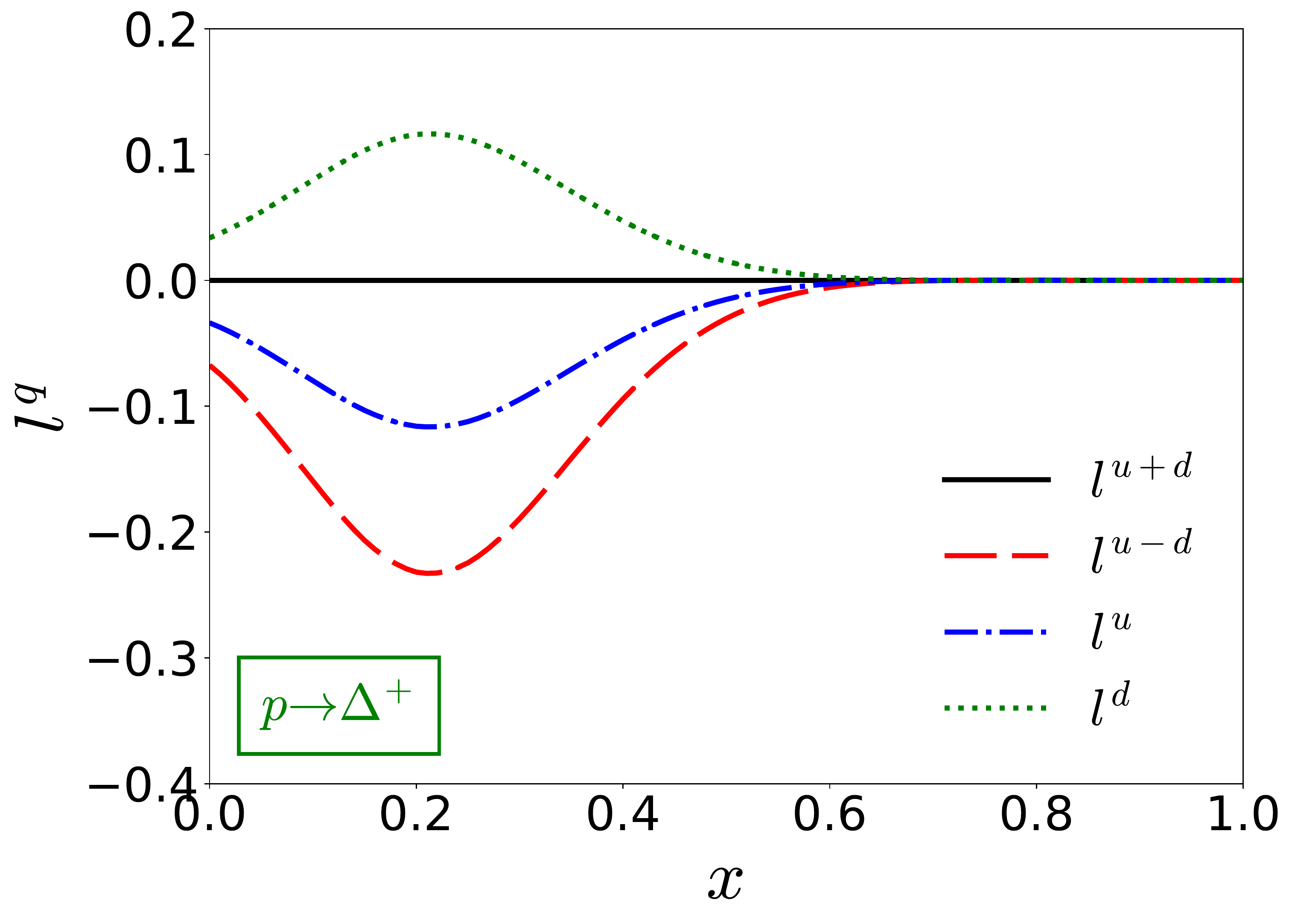}
\caption{OAM quark distribution functions for the proton~(upper left panel), $\Delta^{+}$ baryon~(upper right panel), and $p\to \Delta^{+}$ transition~(lower panel). The solid~(black), dashed~(red), dot-dashed~(blue), and dotted~(green) lines represent the $l^{u+d}$, $l^{u-d}$, $l^{u}$, and $l^{d}$ contributions, respectively.}
\label{fig:2}
\end{figure}
Figure~\ref{fig:2} illustrates the OAM quark distribution functions. We observe that the OAM contribution $l^{q}$ to the baryon spin is relatively small compared to $\Delta q$. This suggests that the nonrelativistic approximation is a reasonable approximation for describing the total AM. The OAM is parameterized in terms of the three quark distributions $\Phi^{L_{1},L_{2},L_{3}}(x)$, and their normalizations $\alpha^{L_{1},L_{2},L_{3}}$ are estimated as follows:
\begin{align}
\int dx \, \Phi^{L_{1}}(x)=\alpha^{L_{1}}=0.050, \quad \int dx \, \Phi^{L_{2},L_{3}}(x)=\alpha^{L_{2},L_{3}}=-0.010, \quad \mathrm{with} \quad \Phi^{L_{2}}(x) = \Phi^{L_{3}}(x).
\end{align}
We then arrive at the value of the isoscalar OAM
\begin{align}
l^{u+d}_{p,\Delta^{+}} = \int dx \, \left[\Phi^{L_{1}}(x)- \Phi^{L_{2}}(x) - \Phi^{L_{3}}(x)\right] = \int dx \, \Phi^{L}(x) = \alpha^{L}=0.070.
\end{align}
This result is in agreement with the numerical estimate made in Ref.~\cite{Lorce:2011ni}. In the nonrelativistic limit, the OAM quark distributions apparently become null
\begin{align}
\Phi^{L_{1},L_{2},L_{3}}_{\mathrm{NR}}(x) = 0.
\end{align}
Since the quark has no relativistic motion in this limit, all observables relevant to the OAM should be zero.

In the case of the proton, the OAM $l^{u+d}$ is similar in magnitude to $l^{u-d}$, indicating that $l^{d}$ is close to zero. This smallness of the $d$-quark contribution arises from the fully dynamical reason, which is the cancellation between the two dynamical parameters $-\frac{1}{3}\alpha^{L_{1}} \sim -0.017 $ and $-\frac{4}{3}\alpha^{L_{2}} \sim 0.013 $.  However, in the $\Delta^{+}$ baryon, $l^{u}_{\Delta^{+}}$ is twice as large as $l^{d}_{\Delta^{+}}$. This relation holds exactly for the intrinsic spin $\Delta q$, specifically $\Delta u_{\Delta^{+}} =2 \Delta d_{\Delta^{+}}$. In addition, it is noteworthy that there is a significant flavor asymmetry of the OAM in the $p\to \Delta$ transition. The normalizations of these quantities can be found in Table~\ref{tab:1}. The equal but opposite contributions of the $u$ and $d$ quarks to the OAM result in $l^{u+d}_{p\to\Delta^{+}}=0$. When the OAM and the intrinsic spin are combined, they give the total AM. As shown in Eq.~\eqref{eq:sumrule}, numerically the total AM is indeed normalized to the baryon spin.
\begin{table}[htp]
\centering
\setlength{\tabcolsep}{3pt}
\renewcommand{\arraystretch}{1.5}
\caption{Intrinsic spin, OAM, and total angular momentum of the nucleon, the $\Delta^+$ baryon, and the $p\to\Delta$ transition are listed for both the nonrelativistic (NR) and relativistic (Rel.) cases.}
\begin{tabular}{ c |  c | c c c  c | c  c c c | c c  c c } 
\hline
\hline
 \multicolumn{2}{c |}{Contents} &  \multicolumn{4}{c |}{$l^{q}$} & \multicolumn{4}{c |}{$\Delta q$} & \multicolumn{4}{c}{$J^{q}$}    \\
 \hline
 \multicolumn{2}{c |}{$q$}  & $u$ & $d$ & $u-d$ & $u+d$  &$u$ & $d$ & $u-d$ & $u+d$ & $u$ & $d$ & $u-d$ & $u+d$    \\
\hline
\multirow{3}{*}{NR} & $p\to p$  & $0$ & $0$ & $0$ & $0$ & $2/3$ & $-1/6$& $5/6$ & $1/2$ & $2/3$ & $-1/6$& $5/6$ & $1/2$\\
 &$\Delta^+\to \Delta^+$  & $0$ & $0$ & $0$  & $0$ & $1/3$ & $1/6$  & $1/6$ & $1/2$ & $1/3$ & $1/6$  & $1/6$ & $1/2$\\
   &$p\to \Delta^+$  & $0$ & $0$ & $0$  & $0$ & $-\sqrt{2}/3$ & $\sqrt{2}/3$& $-2\sqrt{2}/3$ & $0$ & $-\sqrt{2}/3$ & $\sqrt{2}/3$& $-2\sqrt{2}/3$ & $0$\\
\hline
\multirow{3}{*}{Rel.} & $p\to p$  & $0.073$ & $-0.003$ & $0.076$ & $0.070$ & $0.574$ & $-0.144$& $0.718$& $0.431$& $0.647$& $-0.147$& $0.794$ & $0.5$\\
 &$\Delta^+\to \Delta^+$  & $0.046$ & $0.023$ & $0.023$  & $0.070$ & $0.287$ & $0.144$ & $0.144$ & $0.431$ & $0.333$ & $0.167$ & $0.167$ & $0.5$\\
   &$p\to \Delta^+$  & $-0.037$ & $0.037$ & $-0.074$  & $0$ & $-0.406$ & $0.406$ & $-0.812$ & $0$ & $-0.443$ & $0.443$ & $-0.887$ & $0$\\
\hline
\hline
\end{tabular} 
\label{tab:1}
\end{table}

\section{Large $N_{c}$ analysis of the quark distribution functions \label{sec:4}}
Another way to estimate the value of the quark distribution function is to use the spin-flavor structure in the large $N_{c}$ limit of QCD. In practice, this structure can be obtained within the chiral soliton approach. One of the most realistic and representative models of this approach is the $\chi$QSM. In this model the various quark distribution functions have been evaluated~\cite{Diakonov:1997vc, Diakonov:1996sr, Pobylitsa:1998tk, Schweitzer:2001sr, Goeke:2000wv}. From the given quark distribution functions of the nucleon, one can easily map those of the $N\to \Delta$ transitions by using the spin-flavor symmetry. In fact, the results of this approach are more reliable than those of the LC$\chi$QSM. While in the LC$\chi$QSM the infinite tower of higher-fock states is truncated, all sea-quark contributions (quark-antiquark pairs) are explicitly taken into account in their estimation~\cite{Schweitzer:2012hh}. 

Thus, this section is devoted to the extraction of the quark distribution functions for the $N\to \Delta$ transitions from those for the nucleon using the leading order $1/N_{c}$ expansion relations. While the numerical data for the longitudinally polarized quark distribution functions for the nucleon in the $\chi$QSM are given in Ref.~\cite{Diakonov:1997vc, Diakonov:1996sr}, those for the OAM quark distribution are missing. Thus, we will discuss only the longitudinally polarized quark distribution functions.

First we want to mention the $N_{c}$ scalings of the kinematical variables. The baryon masses are of order $M_{N} \sim M_{\Delta} \sim \mathcal{O}(N_{c})$ and their mass splitting is $M_{N}-M_{\Delta} \sim \mathcal{O}(N^{-1}_{c})$. The 3-momenta are $\bm{p},\bm{p}' \sim \mathcal{O}(N^{0}_{c})$ and the 3-momentum transfer is $\bm{\Delta} \sim \mathcal{O}(N^{0}_{c})$. In addition, the $N_{c}$ scalings of the GPDs arguments are $x,\xi \sim \mathcal{O}(N^{-1}_{c})$ and $t\sim \mathcal{O}(N^{0}_{c})$.

In large $N_{c}$ limit, the longitudinally polarized quark distribution functions are derived as follows~\cite{Diakonov:1997vc, Diakonov:1996sr, Goeke:2001tz, Schweitzer:2016jmd}: 
\begin{align}
&2 S^{z}(S'_{3},S_{3}) [\Delta u(x)+ \Delta d(x)]_{B\to B'}  =  \langle S^{3} \rangle_{B'B} \,  \bigg{(}\Delta u_{\mathrm{sol}}(x) + \Delta d_{\mathrm{sol}}(x)\bigg{)} \sim N_{c} \times f( N_{c} x ), \cr
&2 S^{z}(S'_{3},S_{3}) [\Delta u(x)- \Delta d(x)]_{B\to B'}  =  \langle D^{33} \rangle_{B'B} \,  \bigg{(}\Delta u_{\mathrm{sol}}(x) - \Delta d_{\mathrm{sol}}(x)\bigg{)} \sim  N^{2}_{c} \times f( N_{c} x ).
\end{align}
where $\langle ... \rangle_{B'B}$ stands for the matrix element of the spin-flavor operator between the initial~($B= \{S,S_{3},I,I_{3}\}$) and final~($B'= \{S',S'_{3},I',I'_{3}\}$) baryon states:
\begin{align}
\langle S^{3} \rangle_{B'B} &= \sqrt{S(S+1)} C^{S'S'_{3}}_{S S_{3} 1 0} \delta_{S' S} \delta_{I'_{3} I_{3}}\delta_{I' I}, \quad \langle D^{33} \rangle_{B'B} =- \sqrt{\frac{2S+1}{2S'+1}} C^{S'S'_{3}}_{S S_{3} 1 0} C^{I'I'_{3}}_{I I_{3} 1 0}.
\end{align}
Note that $S~(S_{3})$ and $I~(I_{3})$ stand for the spin~(spin projection) and isospin~(isospin projection) quantum numbers, respectively.
The explicit expressions of the quark distributions $\Delta u_{\mathrm{sol}}$ and $\Delta d_{\mathrm{sol}}$ in the chiral soliton approach are given in Refs.~\cite{Diakonov:1997vc, Diakonov:1996sr}. For the isoscalar, the proton and $\Delta^+$ baryon quark distribution functions are equivalent to each other
\begin{align}
[\Delta u (x)+ \Delta d(x)]_{N\to N}=[\Delta u(x)+ \Delta d(x)]_{\Delta\to \Delta}, \quad [\Delta u(x)+ \Delta d(x)]_{N\to \Delta}=0,
\end{align}
and the isoscalar transition is obviously not allowed because the isospins of the $N$ and $\Delta$ baryons differ by $|I'-I|=1$, so the operator must be $|I'-I| \geq 1$. As for the isovector component, the longitudinally polarized quark distribution for $p \to p$ is related to those for $p \to \Delta^{+}$ and $\Delta^{+} \to \Delta^{+}$
\begin{align}
[\Delta u(x)- \Delta d(x)]_{p\to p} = \frac{1}{\sqrt{2}}[\Delta u(x)- \Delta d(x)]_{p\to \Delta^{+}} = 5[\Delta u(x)- \Delta d(x)]_{\Delta^{+} \to \Delta^{+}}.
\end{align}
Integrating these quark distribution functions over $x$, one obtains the axial-charge relations
\begin{align}
g^{u-d}_{A, p \to p} = \frac{1}{\sqrt{2}}  g^{u-d}_{A, p \to \Delta^{+}}= 5  g^{u-d}_{A, \Delta^{+} \to \Delta^{+}},
\end{align}
which is exactly the same as the total AM~\cite{Kim:2023xvw}. In that paper~\cite{Kim:2023xvw} it was emphasized that the spin-flavor symmetry in the large $N_{c}$ is blind to the decomposition between the intrinsic spin and the OAM. This is the reason why the spin-flavor relation for the total AM also holds for the isovector axial charge.

In addition, we want to mention the relation of the transition GPDs to the quark distribution functions. The $N\to \Delta$ GPDs~\cite{Goeke:2001tz} are defined as
\begin{align}
&\int \frac{d\lambda}{2\pi} e^{i \lambda x} \langle \Delta^{+} (p',S'_{3}) | \bar{\psi}(-z /2) \slashed{n} \gamma^{5} \tau^{3} \psi(z /2)  | p (p',S'_{3}) \rangle = \bar{u}^{\beta}(p', S'_{3}) \bigg{[}C_{1}(x,\xi,t) n_{\beta} + ... \bigg{]} u (p, S_{3}),
\end{align}
where ... denotes the GPDs suppressed in the large $N_{c}$ limit and dropped in the forward limit $\Delta,\xi =0$, and $\bar{u}^{\beta}$ and $u$ are Rarita–Schwinger and Dirac spinors, respectively. Here $n$ denotes the light-cone vector, and the space-time 4-vector $z^{\mu}$ in Eq.~\eqref{eq:def_pdf} is rewritten as $z^{\mu}= n^{\mu}\lambda$. The first Mellin moment of the transition GPD is related to the axial-vector form factor
\begin{align}
\int dx \, C_{1}(x,\xi, t) =  2 C_{5}(t),
\end{align}
where the matrix element of the local axial vector current is parametrized in terms of the Adler-type form factors~\cite{osti_4806262, PhysRevD.12.2644} as 
\begin{align}
& \langle \Delta^{+} (p',S'_{3}) | \bar{\psi}(0) \gamma_{\mu}\gamma^{5} \frac{\tau^{3}}{2} \psi(0)  | p (p',S'_{3}) \rangle = \bar{u}^{\beta}(p', S'_{3}) \bigg{[}C_{5}(t) g_{\mu \beta} + ... \bigg{]} u (p, S_{3}).
\end{align}
In the forward limit $\Delta,\xi =0$, the transition GPDs are reduced to quark distribution functions in the $N\to \Delta$ transition
\begin{align}
[\Delta u(x) - \Delta d(x)]_{p\to \Delta^{+}} = \sqrt{\frac{2}{3}} C_{1}(x,0,0).
\end{align}
Integrating the quark distribution functions over $x$, one arrives at the axial charge
\begin{align}
g^{u-d}_{A, p\to \Delta^{+}} = \sqrt{2} g^{u-d}_{A, p\to p}  = 2\sqrt{\frac{2}{3}} C_{5}(0)\sim \mathcal{O}(N_{c}).
\end{align}
They coincide with the large $N_{c}$ relations between the nucleon and the $N\to\Delta$ transition GPDs and their $N_{c}$ scalings
\begin{align}
&C_{1}(x,\xi,t) = \sqrt{3} \tilde{H}^{u-d}(x,\xi,t) \sim N^{2}_{c}  \times f(N_{c}x,N_{c}\xi,t).
\end{align}
It is easy to find these large $N_{c}$ relations among the GPDs, the PDFs, and the charge for the $\Delta$ baryon~\cite{Fu:2022bpf, Fu:2023dea}.

Finally, it is worth noting that the same spin-flavor structure holds for the OAM quark distributions
\begin{align}
l^{u-d}_{p\to p}(x) = \frac{1}{\sqrt{2}}l^{u-d}_{p\to \Delta^{+}}(x) = 5l^{u-d}_{\Delta^{+} \to \Delta^{+}}(x),
\end{align}
because the spin-flavor structures of the intrinsic spin, OAM, and total AM are shared.

\begin{figure}[htp]
\includegraphics[scale=0.28]{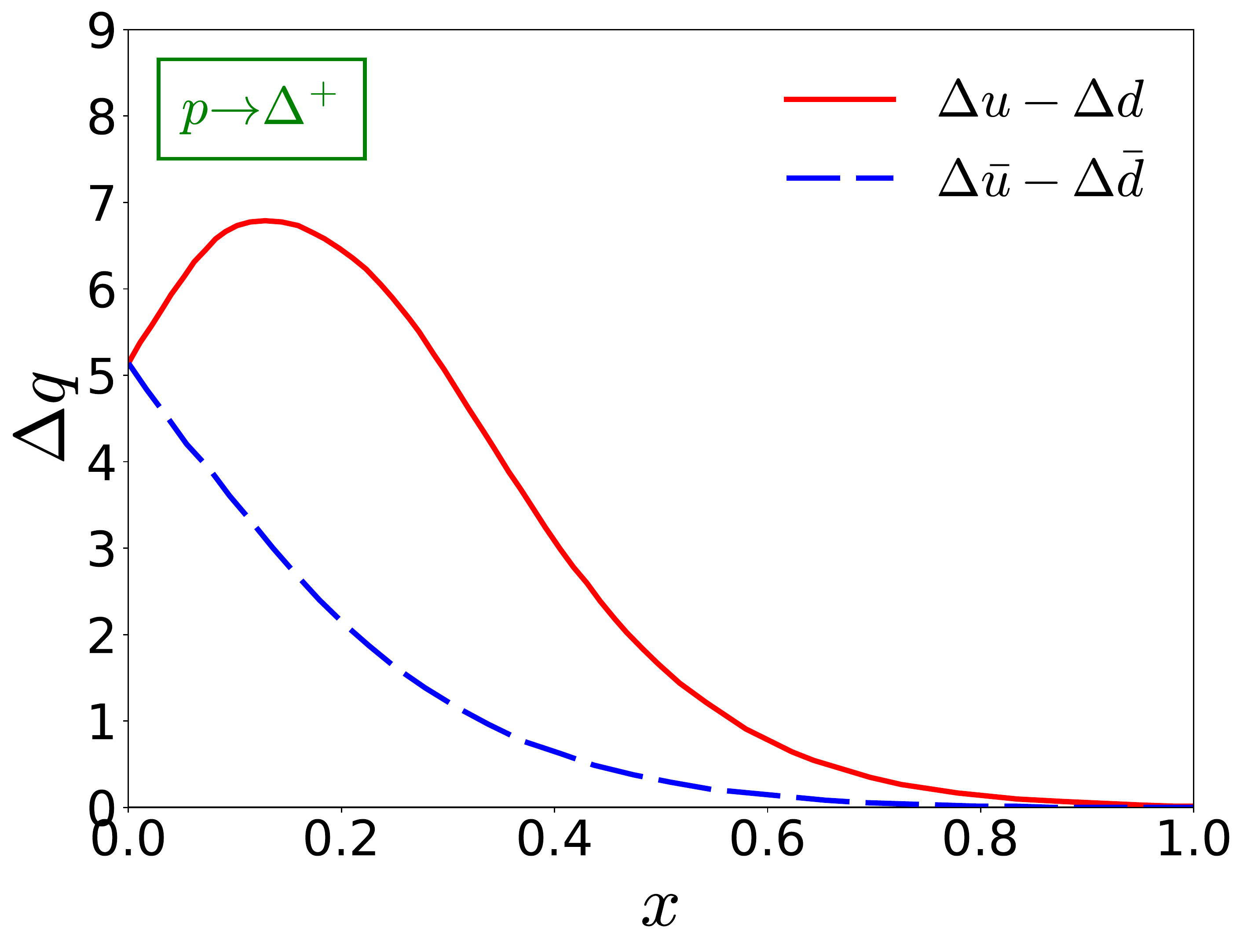}
\includegraphics[scale=0.28]{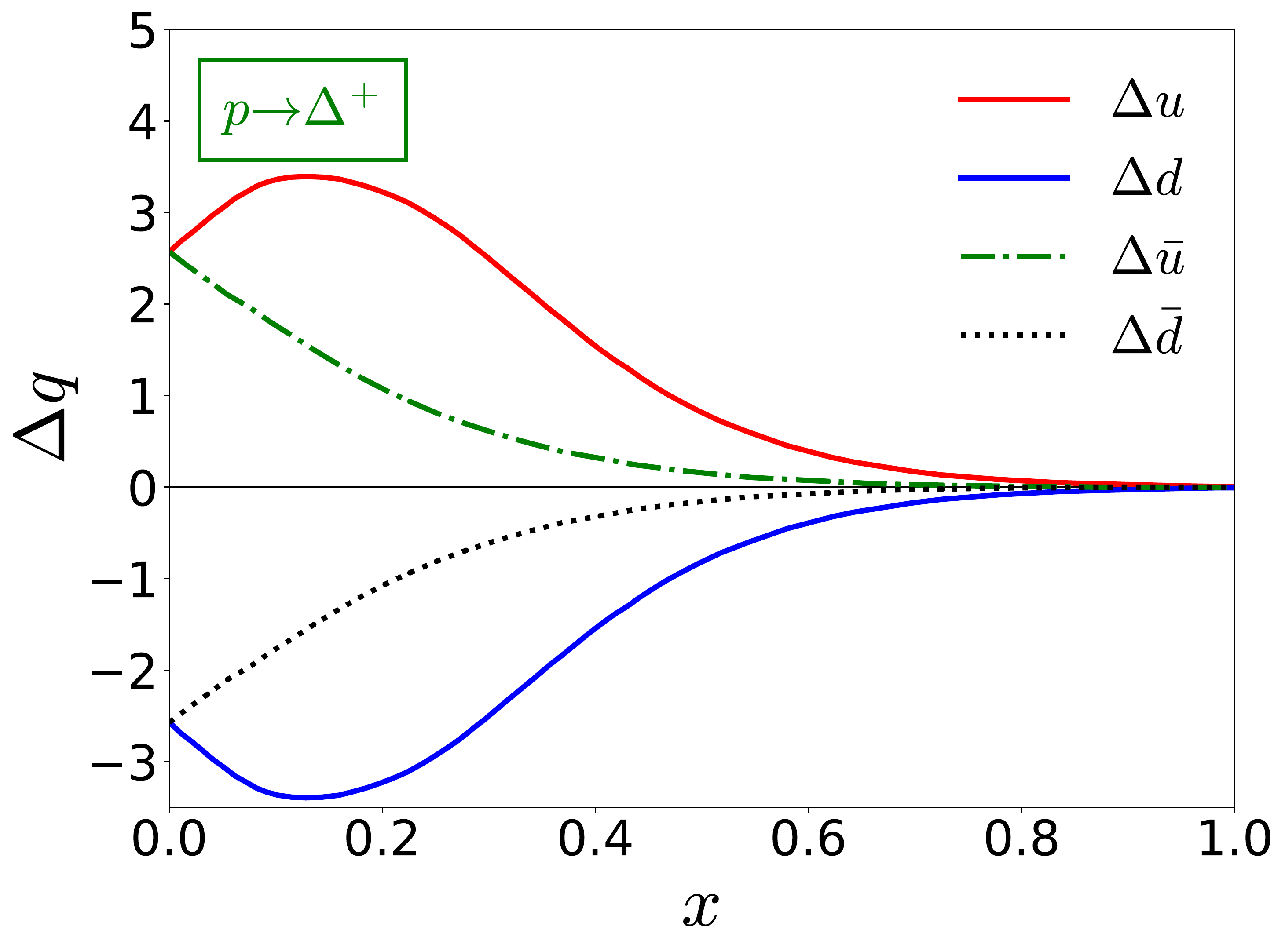}
\includegraphics[scale=0.28]{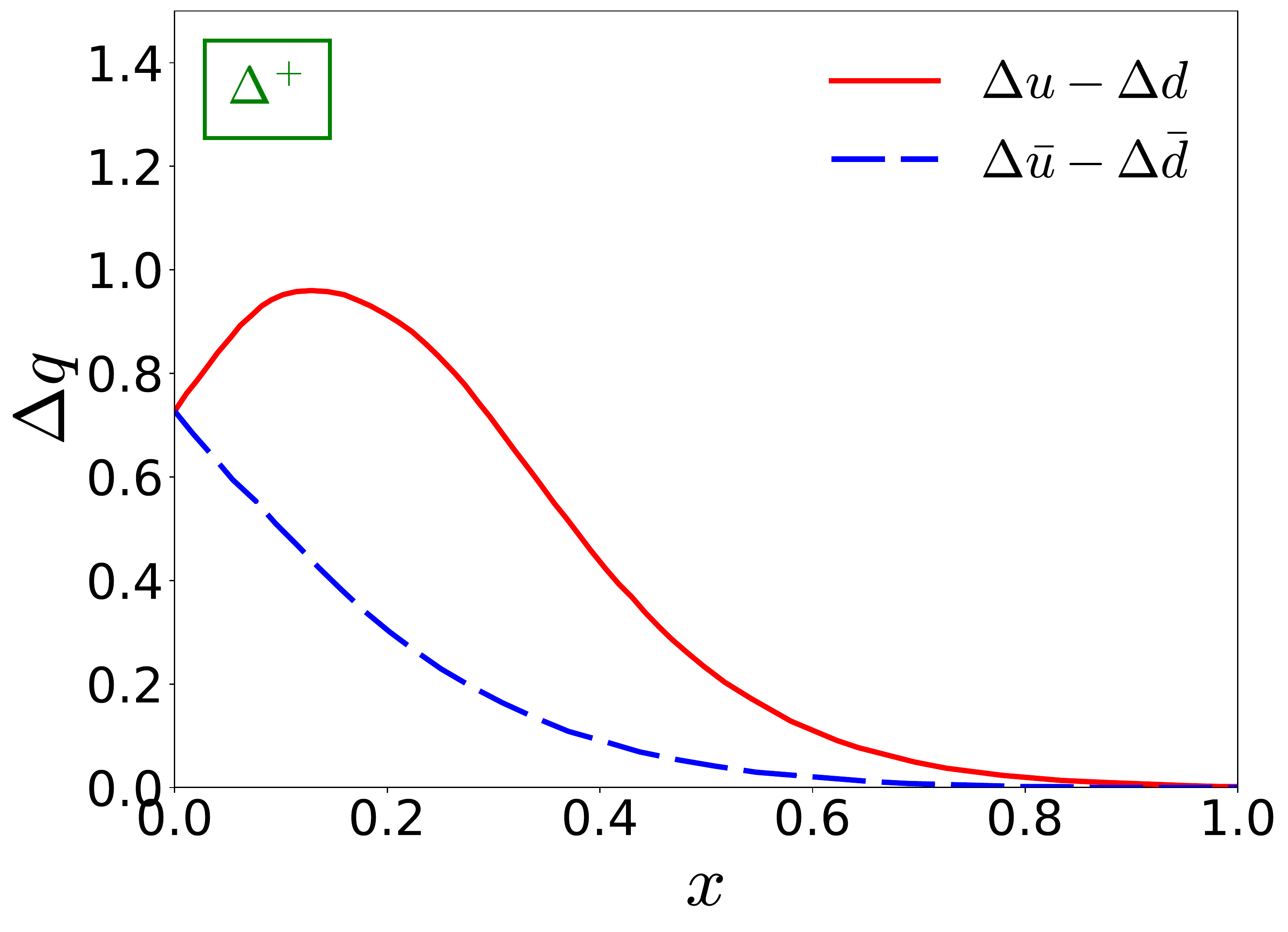}
\caption{Longitudinally polarized quark distribution functions for the $p\to \Delta^{+}$~(upper panels), $\Delta^{+}$ baryon~(lower panel).}
\label{fig:3}
\end{figure}

Figure~\ref{fig:3} shows the longitudinally polarized quark distributions for the $p\to \Delta^{+}$ and $\Delta^{+}$ baryons, based on their spin-flavor structures. The upper left panel shows that the quark and antiquark distribution functions for the $N\to \Delta$ transition are about 1.4 times larger than those of the nucleon. Note that in the large $N_{c}$ limit, the scaling behavior of $\Delta u+ \Delta d$ and $\Delta u- \Delta d$ is of the order of $\mathcal{O}(N^{0}_{c})$ and $\mathcal{O}(N^{1}_{c})$, respectively. Therefore, a mere consideration of the leading contributions is not sufficient for the flavor decomposition, since the subleading contribution to the isovector quark distribution functions $[\Delta u- \Delta d \, |_{\mathrm{sub}} \sim \mathcal{O}(N^{0}_{c})$ must also be taken into account. In contrast, for the $N \to \Delta$ transition, the isoscalar quark distribution functions must be zero, eliminating the $N_{c}$ scaling admixture. Consequently, the flavor-decomposed quark distributions are shown in the right panel of Fig.~\ref{fig:3}. Additionally, for completeness, we include the quark distribution functions for the $\Delta$ baryon. Similar to the LC$\chi$QSM, these distributions are relatively small compared to the nucleon results due to the kinematical factors.

Table~\ref{tab:2} shows the separate contributions of the OAM and the intrinsic spin to the total AM, using data from lattice QCD at $\mu^{2}= 4~\mathrm{GeV}^{2}$ and the $\chi$QSM at $\mu^{2} \sim 0.36~\mathrm{GeV}^{2}$. Instead of relying on the normalization values of the quark distribution functions in Fig.\ref{fig:3}, we use values obtained from calculations of the EMT form factors~\cite{Won:2022cyy, Won:2023rec} and the axial-vector form factors~\cite{Christov:1995vm}. In particular, the lattice QCD predict a substantial flavor asymmetry in both OAM and intrinsic spin. These contributions have opposite signs, leading to a partial cancellation and resulting in a relatively small value of $J^{u-d}$. 
\begin{table}[htp]
\centering
\setlength{\tabcolsep}{3pt}
\renewcommand{\arraystretch}{1.5}
\caption{The intrinsic spin, OAM and total angular momentum of the nucleon, the $\Delta^+$ baryon and the $p\to\Delta$ transition are listed using the input data from the QCD lattice~\cite{LHPC:2010jcs} and the $\chi$QSM~\cite{Won:2022cyy, Won:2023rec, Christov:1995vm}. Input values are marked with an asterisk~$^*$.}
\begin{tabular}{ c |  c | c c | c  c | c  c  } 
\hline
\hline
 \multicolumn{2}{c |}{Contents} &  \multicolumn{2}{c |}{$l^{q}$} & \multicolumn{2}{c |}{$\Delta q$} & \multicolumn{2}{c }{$J^{q}$}    \\
 \hline
 \multicolumn{2}{c |}{$q$}  & $u-d$ & $u+d$ & $u-d$ & $u+d$  & $u-d$ & $u+d$    \\
\hline
\multirow{3}{*}{Lattice QCD~\cite{LHPC:2010jcs} \ $\mu^{2}= 4~\mathrm{GeV}^{2}$ } & $p\to p$   & $-0.38^{*}$ & $0.03^{*}$ & $0.61^{*}$& $0.21^{*}$ & $0.23^{*}$ & $0.24^{*}$\\
 &$\Delta^+\to \Delta^+$   & $-0.08$ & $0.03$ & $0.12$ & $0.21$ & $0.05 $ & $0.24$\\
   &$p\to \Delta^+$  & $-0.54$ & $0$ & $0.86$ & $0$ & $0.33$ & $0$\\
\hline
\multirow{3}{*}{$\chi$QSM~\cite{Won:2022cyy, Won:2023rec, Christov:1995vm} \ $\mu^{2} \sim 0.36~\mathrm{GeV}^{2}$} & $p\to p$   & $-$ & $-$ & $0.61^{*}$& $0.19^{*}$ & $0.56^{*}$ & $0.50^{*}$\\
 &$\Delta^+\to \Delta^+$   & $-$ & $-$ & $0.12$ & $0.19$ & $0.11$ & $0.50$\\
   &$p\to \Delta^+$  & $-$ & $-$ & $0.86$ & $0$ & $0.79$ & $0$\\
\hline
\hline
\end{tabular} 
\label{tab:2}
\end{table}

Here we do not present the isoscalar and isovector OAMs obtained in the $\chi$QSM. This is due to two ambiguities. First, the effective action in the $\chi$QSM can be split into a real part and an imaginary part. The real part of the effective action exhibits ultraviolet (UV) divergences, while the imaginary part is free of such divergences; see Ref.~\cite{Christov:1995vm} for details. In this context, the total isovector AM needs no regularization. This is easily demonstrated by the gradient expansion (i.e., expansion of the quark propagator in terms of the derivatives of the pion field), where the leading order is free of UV divergences. However, the individual contributions of the intrinsic spin and the OAM suffer from UV divergences. Remarkably, the UV divergent parts of OAM and intrinsic spin exactly cancel each other out, resulting in a total AM that is free of UV divergences~\cite{gradient}. Consequently, the separation of OAM and intrinsic spin at the level of the EMT would be inappropriate. Only the total AM $J^{u-d}$ can be considered as a reliable observable when starting from the EMT. On the other hand, since $J^{u+d}$ requires regularization, it would be safer to study the separate contributions of OAM and intrinsic spin. This can be easily demonstrated again using the gradient expansion. This analysis states that at the leading order of the gradient expansion~\cite{gradient} all OAM carry the nucleon spin; see also Ref.~\cite{Granados:2019zjw} in the context of chiral effective field theory. Second, in a chiral theory, the decomposition of the isovector part of the Ji's AM into the spin and OAM is spoiled due to a interacting term; see Refs.~\cite{Wakamatsu:2005vk, Wakamatsu:2006dy}. In addition, the QCD relation for the chiral-odd twist-3 quark distributions is also violated~\cite{Schweitzer:2003uy, Ohnishi:2003mf}. This may be due to the ambiguity of the identification of the twist-3 QCD operator with the effective operators.

In our predictions in the large $N_{c}$ limit, we determine the separate contributions of intrinsic spin and OAM for both the $\Delta$ baryon and the $N\to \Delta$ transition. Remarkably, the results show significant magnitudes for the isovector components of both OAM and intrinsic spin. However, in the case of the $\Delta$ baryon, the magnitudes of the isovector components for OAM and intrinsic spin are comparatively smaller, in agreement with the expectations of the LC$\chi$QSM.

\section{Summary and conclusions \label{sec:5}}
The purpose of this study is to investigate the quark distribution functions associated with the $N\to \Delta$ transition and to address a dynamical aspect regarding the decomposition of the contributions from orbital angular momentum and intrinsic spin.

Starting from the energy-momentum tensor current, we establish the definition of angular-momentum quark distribution functions applicable to all initial and final baryon states, including the specific case of the $N\to \Delta$ transition. These quark distribution functions include the longitudinally polarized quark distribution and the orbital angular-momentum quark distribution. They can be related to the twist-2 and twist-3 generalized parton distributions, respectively.

To estimate the angular-momentum quark distribution functions, we use two different approaches. The first approach involves the use of overlap representations of the $3Q$ light-cone wave function derived from the chiral quark-soliton model. The second approach relies on the standard spin-flavor symmetry in the framework of the large $N_{c}$ limit of QCD.

First, the determination of the orbital angular momentum quark distribution function involves two independent distributions, $\Phi^{L_{1}}(x)$ and $\Phi^{L_{2}}(x)$, derived from the overlap representation of the $3Q$ light cone wave function. It turns out that a linear combination of the proton and neutron orbital angular momentum quark distributions allows the extraction of those associated with the $N\to \Delta$ transition and the $\Delta$ baryon. On the other hand, the longitudinally polarized quark distribution relies solely on the single distribution $\Phi^{A}(x)$. Thus, access to the quark distributions of the proton allows the determination of those for the $N\to\Delta$ transition and the $\Delta$ baryon.

We found that a significant fraction of the QCD angular momentum comes from the intrinsic spin, while the remaining fraction is due to the relativistic motion of the quarks, known as orbital angular momentum. As a result, the nonrelativistic approximation holds reasonably well for the $3Q$ light-cone wave function. Nevertheless, both the intrinsic spin and the orbital angular momentum show a significant flavor asymmetry. This suggests that the isovector component of the total angular momentum is substantial. The substantial flavor asymmetry in the angular momentum holds for both the $p\to p$ and $p\to \Delta^{+}$ processes. For the $\Delta^{+}$ baryon, however, the isovector component is suppressed due to the associated kinematical factor.

Second, we use the spin-flavor symmetry in the large $N_{c}$ limit of QCD to derive the longitudinally polarized quark distribution functions for the $\Delta^{+}$ baryon and the $p \to \Delta^{+}$ transition. Using the dynamical information obtained in the nucleon sector, we can easily extend the quark distribution functions to different baryon quantum numbers. To achieve this, we use data from the chiral quark-soliton model, which provides reliable quark distribution functions for the nucleon. A notable advantage of this approach is that it explicitly includes an infinite number of quark-antiquark pair contributions, in contrast to a truncated $3Q$ light-cone wave function. Furthermore, we establish the relation between the generalized quark distributions and the quark distribution functions for the $N\to \Delta$ transition. It is important to emphasize that the spin-flavor relation observed in the longitudinally polarized quark distribution functions is also applicable to functions associated with orbital angular momentum and total angular momentum. This is because the spin-flavor symmetry does not distinguish between spin, orbital angular momentum, and total spin.

Using data from lattice QCD and the chiral quark-soliton model for the nucleon, we make predictions about the orbital angular momentum, intrinsic spin, and total angular momentum of the $\Delta^{+}$ baryon and the $p \to \Delta^{+}$ transition. While the isoscalar orbital angular momentum, intrinsic spin, and total angular momentum for the $p \to \Delta^{+}$ transition become zero due to the isospin properties, substantial flavor asymmetries are observed. Interestingly, most of these asymmetries cancel out due to the opposite signs of the orbital angular momentum and the intrinsic spin. Consequently, the total angular momentum is relatively small compared to the individual contributions of orbital and intrinsic spin. Nevertheless, we find a significant flavor asymmetry in the isovector component of the total angular momentum, which is consistent with the expected size in the large $N_{c}$ limit of QCD. Moreover, our results for $\Delta^{+}$ baryon are consistent with those obtained from the overlap representation of the $3Q$ light-cone wave function, indicating that the overall smallness of the intrinsic spin, orbital angular momentum, and total angular momentum for the $\Delta$ baryon can be attributed to the kinematical factor.

It would be interesting to study the orbital angular momentum quark distribution functions for the nucleon in the framework of the chiral quark-soliton model. In addition, it would be interesting to study the parity-odd partner of the energy-momentum tensor. It contains rich information about the partonic structure of the nucleon, such as the spin-orbit correlation and the second moments of the quark helicity distribution~\cite{Lorce:2014mxa}.

\section{acknowledgement}
The author is indebted to Ho-Yeon Won for valuable discussions and for providing numerical data from the $\chi$QSM, and to C. Weiss, and J. L. Goity, and H.-Ch. Kim for their help and advice. The author also wants to express gratitude to J.-W. Qiu and C. Weiss for their careful reading and comments. This work is supported by the U.S. Department of Energy, Office of Science, Office of Nuclear Physics under contract DE-AC05-06OR23177, and also the author acknowledge partial support by the U.S. Department of Energy, Office of Science, Office of Nuclear Physics under the umbrella of the Quark-Gluon Tomography (QGT) Topical Collaboration with Award DE-SC0023646.

\newpage
\bibliographystyle{apsrev4-2}
\bibliography{LCWF_AM}
\end{document}